# Divergent phonon angular momentum driven by temperature and strain

Young-Jae Choi and Seung-Hoon Jhi*

*Physics Department, POSTECH, Cheongam-ro 77, Pohang 37673, South Korea*

*Corresponding author: jhish@postech.ac.kr

**ABSTRACT**

The phonon angular momentum (PAM) may exhibit exotic temperature dependence as it is sensitive to the phonon lifetime. Constant phonon-lifetime approximation fails to depict such behavior. Here, we study the PAM of AlN, GaN, and graphene-like boron nitride (g-BN) monolayer with full consideration of phonon lifetime using first-principles calculations. We show that wurtzite AlN and GaN acquire divergent PAM at low temperatures from their lowest-lying phonon branches. The g-BN monolayer, on the other hand, does not have finite PAM at equilibrium structure. Rather it shows intriguing strain-dependence in PAM; the compressive strain greater than the critical size generates divergent PAM at low temperatures due to the divergent lifetime of TA phonons. As PAM couples with rotational excitations in solids associated with charge, spin, or electromagnetic fields, our study demonstrates a possibility of mechanical and thermal engineering of such excitations.

## I. INTRODUCTION

The orbital motion of atoms may generate finite angular momentum in solids [1]. Namely, the phonon angular momentum (PAM) can exist in a sample without inversion or time-reversal symmetry and can be exchanged with other angular momentum degrees of freedom such as electron spin, orbital, and rigid body angular momenta [1-3]. The existence of PAM was experimentally proven by helicity-resolved Raman scattering measurement on four kinds of two-dimensional layered transition-metal dichalcogenides [4]. The chiral phonons that mediate the intervalley hole transfer in $WSe_2$ were detected using infrared circular dichroism measurement [5]. The spin magnetic moment of a Ni film was shown to be directly controlled by the injection of PAM [6]. The theory of dynamical multiferroicity, *i.e.*, the induction of the phonon magnetic moment by the time-dependent electric polarization of chiral phonons in an external field, has been discussed [7-9]. The splitting of phonon modes under high magnetic field in experiment ("phonon Zeeman effect") was interpreted as a proof of the phonon magnetic moment [10, 11]. Some theoretical and numerical studies have also investigated the PAM of planar materials. The PAM of graphene-like boron nitride (g-BN) was described in detail from the atomic vibration to its chirality [12, 13], and the heterostructure of graphene and g-BN was found to carry chiral phonons in its valleys [14]. Also, the gate-tunable circular phonon dichroism of strained graphene was reported [15].

An interesting notion of PAM is its link to the thermal Hall effect. The thermal Hall conduction can arise from various carriers such as electrons, magnons [16], spinons [17], phonons [18], and Majorana edge modes [19], yet its microscopic understanding in real materials is very desirable. Lanthanum cuprate in its pseudogap phase and paramagnetic $SrTiO_3$ were reported to exhibit large thermal Hall conduction, which is attributed to chiral phonons [20, 21] or linked to the phonon Hall effect [22, 23]. The concept of PAM originally forked from a study on the phonon Hall effect [1]. The angular motion of nuclei can interact with magnetic elements, which may lead to the phonon Hall effect and topologically nontrivial phonon bands [24-26]. The Raman-interaction strength approximately proportional to the magnitude of PAM [24-26] indicates that the materials with large PAM are likely to produce large phonon Hall conduction. Drastic increase of PAM thus has significant implication with regard to the thermal Hall effect and related phenomena.

When a temperature gradient is applied to a finite-size sample that does not have the inversion symmetry, it can acquire non-zero total PAM, namely the phonon Edelstein effect, and exhibit the phonon Einstein-de Haas effect [3, 27]. If a solid sample possesses finite PAM and Born-effective charges simultaneously, its phonon generates a magnetic moment [27, 28]. Polar materials, which can possess PAM even without external fields, have a possibility for unique phenomena such as the anomalous phonon Hall effect.

The theory of the PAM has been developed within the approximation of constant lifetime, which is in fact only valid at high temperatures above the Debye temperature [3, 27]. However, the phonon lifetime may diverge at low temperatures due to the restrictions on phonon-phonon interaction [29] and as a result, PAM may change drastically at low temperatures. Also, external strains can change the crystal symmetry, and thereby open the possibility of controlling PAM. Here, we investigate the PAM of III-V polar materials in two and three dimensions when the temperature is lowered below or an external strain is applied above a threshold level that restricts the phonon decay channels.

## II. METHODS

The angular momentum of an atom is calculated as a vector product of the displacement and momentum vectors, and the total atomic angular momentum is obtained by summing all atomic angular momenta in the system [1]. The displacements and momenta of atoms in a solid can be expressed in terms of phonon modes, and so does the PAM. We note that the PAM here is termed more specifically as spin angular momentum of phonons in the literatures [30]. Total PAM per unit volume or area is then obtained by Fourier transforming the displacement vector to crystal momentum,

$$J_i = \frac{1}{V} \sum_{s=(\mathbf{k},\sigma)} l_{s,i} \left[ f_0(\omega_s) + \frac{1}{2} \right], \qquad (1)$$

where $V$ is the volume or area of the sample, $\mathbf{k}$ is the crystal momentum, $\sigma$ is the phonon branch index, $l_{s,i}$ is the mode-specific PAM, $f_0$ is the Bose-Einstein distribution, and $\omega_s$ is phonon frequency [1]. The mode-specific PAM, i.e., the angular momentum of a single phonon mode, is defined as $l_{\sigma,i}(\mathbf{k}) = \hbar \boldsymbol{\epsilon}_\sigma^\dagger(\mathbf{k}) \mathbf{M}_i \boldsymbol{\epsilon}_\sigma(\mathbf{k})$, where $M_{ijk} = I_{n \times n} \otimes (-i)\varepsilon_{ijk}$, $\boldsymbol{\epsilon}_\sigma(\mathbf{k})$ is a phonon eigenvector, $I$ is the identity matrix, and $\varepsilon_{ijk}$ is the Levi-Civita symbol [1]. The mode-specific PAM has an upper limit of $\hbar$ in magnitude since the eigenvectors are normalized and the matrix $M_{ijk}$ is unitary.

The mode-specific PAM vector $\mathbf{l}_\sigma(\mathbf{k})$ is always zero unless the inversion or time-reversal symmetry is broken [3, 27]. For a sample with the time-reversal symmetry, $\mathbf{l}_\sigma(\mathbf{k})$ is an odd function of the crystal momentum, so total PAM is zero in equilibrium even without the inversion symmetry [27]. However, a sample in a non-equilibrium condition such as in the presence of a temperature gradient along a certain direction, can have a non-zero total PAM [27]. In such a case, total PAM is derived from the deviation of $f_0$ in Eq. 1 up to the next leading term based on the Boltzmann transport theory and Bhatnagar, Gross, and Krook approximation [31],

$$J_i = -\frac{1}{V} \sum_s l_{s,i} \tau_s \mathbf{v}_s \cdot \nabla T \frac{\partial f_0}{\partial T} = \sum_j \alpha_{ij} \frac{\partial T}{\partial x_j}, \qquad (2)$$

where $\tau_s$ is phonon lifetime, $\mathbf{v}_s$ is phonon group velocity, and $\alpha_{ij}$ is the PAM response tensor [27]. We assume a small temperature gradient, so the system is not far from equilibrium. The temperature-dependent terms in $\alpha_{ij}$ are $\tau_s$ and $\partial f_0/\partial T$. The phonon lifetime is calculated from the linewidth $\tau_s = (2\Gamma_s)^{-1}$, which is the imaginary part of an anharmonic self-energy $\Sigma_s(\omega) = \Delta_s(\omega) - i\Gamma_s(\omega)$ of phonon bubble diagram [32]. The explicit form of the linewidth is given as

$$\Gamma_s(\omega) = \frac{18\pi}{\hbar^2} \sum_{s_1 s_2} |\Phi_{-ss_1 s_2}|^2 \{(f_{s_1} + f_{s_2} + 1)[\delta(\omega - \omega_{s_1} - \omega_{s_2}) - \delta(\omega + \omega_{s_1} + \omega_{s_2})] \\ + (f_{s_1} - f_{s_2})[\delta(\omega + \omega_{s_1} - \omega_{s_2}) - \delta(\omega - \omega_{s_1} + \omega_{s_2})]\}, \quad (3)$$

where $\Phi$ is the third-order force constant matrix [32]. The higher-order phonon interactions are not considered in this study.

Polar materials with an internal electric field like wurtzite have finite PAM texture [27]. When a temperature gradient is applied in a direction perpendicular to the polarization, the phonon distribution becomes unbalanced across the sample and heat transport is generated to reestablish the equilibrium [27]. In such a nonequilibrium state, a finite total PAM occurs in the direction perpendicular to both polarization and heat current (Eq. 2) [27]. In return, the sample gains the rigid-body angular momentum known as the phonon Einstein-de Haas effect [27].

The total energies and forces were calculated using DFT code implemented in the Vienna Ab initio Simulation Package (VASP) [33] with the projector augmented wave method [34]. The electron exchange-correlation was treated within the generalized gradient approximation in the form of the Perdew-Burke-Ernzerhof functional [35]. The second-order force constants were calculated by VASP and Phonopy codes and corrected with translational and rotational sum-rules implemented by the Hiphive package [36, 37]. All dynamical matrices calculated by first-principles calculations were corrected using a non-analytical term correction implemented in the Phonopy code [38].

Wurtzite AlN and GaN of our calculations have hexagonal lattice symmetry with the lattice parameters $a_1^{\text{AlN}}$=3.112 Å, $a_3^{\text{AlN}}$=4.983 Å, $a_1^{\text{GaN}}$=3.231 Å, and $a_3^{\text{AlN}}$=5.255 Å. For the structure optimization and Born effective charge calculations, the k-grids were set to 17×17×10 for AlN and 17×17×9 for GaN. The calculated Born effective charges are $Z^*_{\text{Al},xx}$=2.513$e$, $Z^*_{\text{Al},zz}$=2.674$e$, $Z^*_{\text{Ga},xx}$=2.687$e$, and $Z^*_{\text{Ga},zz}$=2.825$e$. The finite displacement method implemented in the Phonopy code was used for the first-principles harmonic phonon calculations [38]. The 4×4×2 supercells of the primitive unit cell and k-grids of 5×5×5 for electronic structures were used to calculate the harmonic force constants of both AlN and GaN. The Phono3py code was used to calculate the third-order force constants and three-phonon processes [32]. The 3×3×2 supercell and 6×6×5 k-grids were used to calculate the third-order force constants.

Monolayer g-BN has 2-dimensional graphene-like hexagonal structure. The structure parameter was optimized to $a^{\text{BN}}$=2.517 Å with 30 Å-thick vacuum buffers between layers along the $z$-axis. For the structure optimization and Born effective charge calculations, the k-grids were set to 18×18×1. The Born effective charge was calculated as $Z^*_{\text{B},xx}$=2.708$e$. The 4×4×1 supercells and 5×5×1 k-grids were used for calculation of the harmonic force constants. The 3×3×1 supercells and 7×7×1 k-grids were used for calculation of the third-order force constants of g-BN.

### III. RESULTS

#### A. Wurtzite AlN and GaN

Wurtzite AlN and GaN have $C_{3v}$ point group symmetry, and their mode-specific PAM is always perpendicular to the $z$-axis. The mode-specific PAM in the wurtzite has a chiral texture in momentum space that consists mainly of the tangential component (Fig. 1).

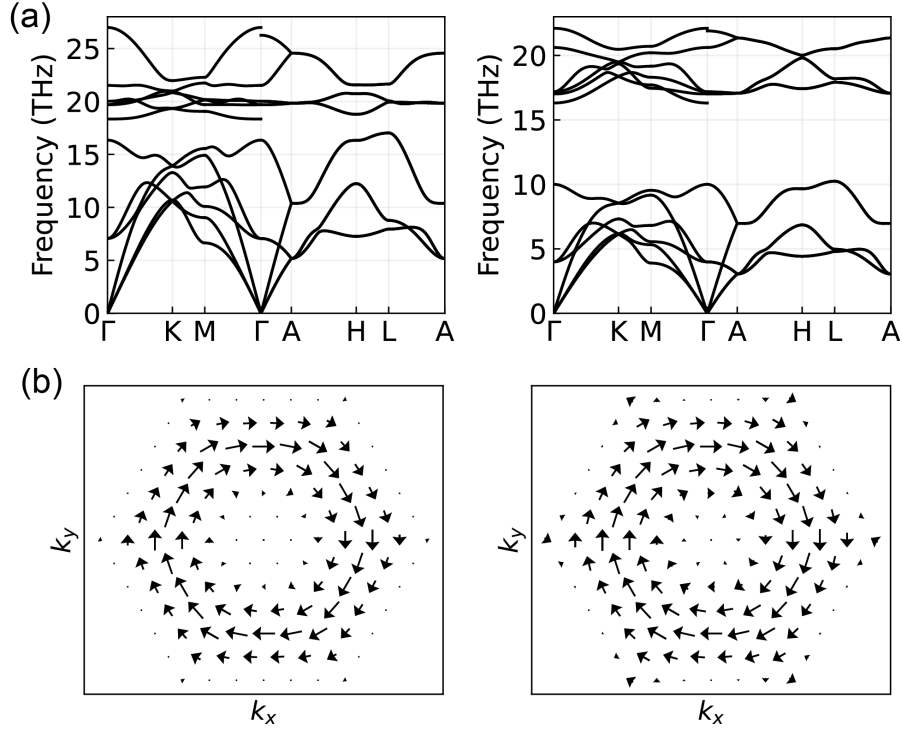

**Fig. 1. (a) Phonon band structures of AlN and GaN and (b) mode-specific PAM texture of the sixth branch.** Left panel, AlN; right panel, GaN.

A single independent component in the PAM response tensor of the wurtzite is $\alpha_{xy}$. The $\alpha_{xy}$ plot of AlN and GaN (Fig. 2) indicates that the PAM response tensor depends heavily on the phonon lifetime, especially at temperature < 100 K. Without the lifetime dependence, $\alpha_{xy}$ went to zero as the temperature was lowered (Fig. 2a). However, when the phonon lifetime was considered correctly, $\alpha_{xy}$ diverged greatly as temperature approaches zero (Fig. 2b). The magnitude of $\alpha_{xy}$ increased by about 40-100 times at 30 K from that at room temperature (from $8.61 \times 10^{-17}$ to $-3.43 \times 10^{-15}$ Js/m$^2$K for AlN and from $3.94 \times 10^{-16}$ to $-5.52 \times 10^{-14}$ Js/m$^2$K for GaN). Moreover, the crossover in the sign of PAM at 30-40 K cannot be captured using the constant lifetime approximation.

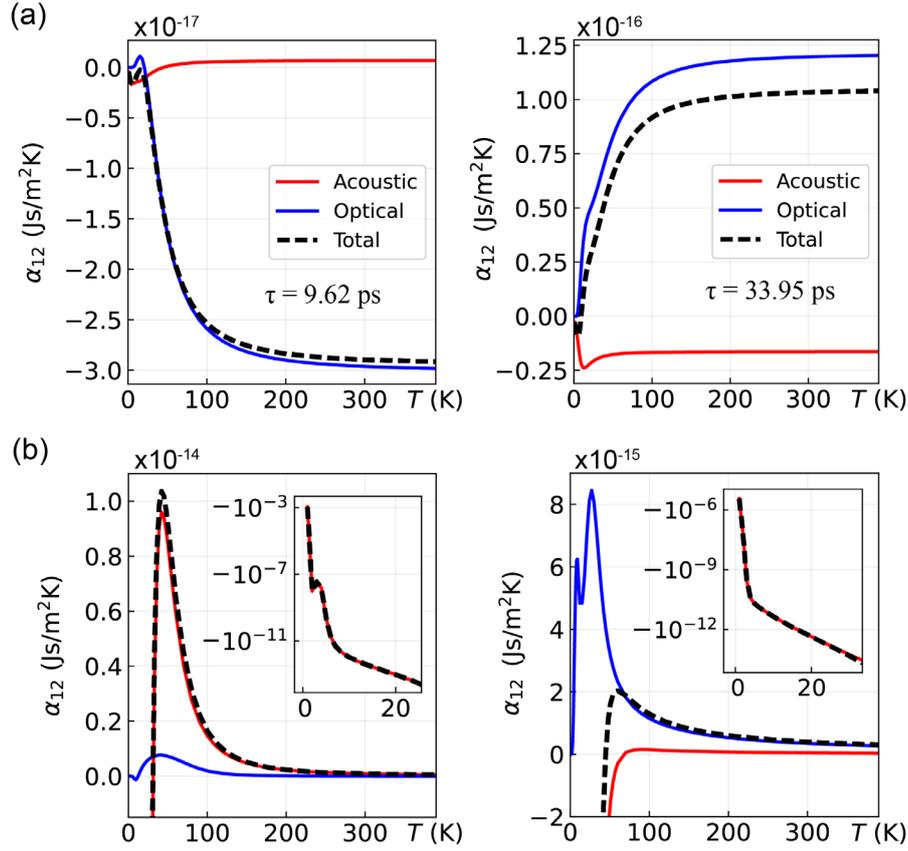

**Fig. 2. PAM response tensor $\alpha_{xy}$ of AlN (left panel) and GaN (right panel).** (a) Constant phonon lifetime ($\tau$= 9.62 and 33.95 ps) and (b) mode-specific phonon lifetime are used. The contributions of acoustic (red) and optical modes (blue), and the total (black dashed) are plotted. Insets in (b) used logarithmic scale.

This drastic behavior of $\alpha_{xy}$ at low temperatures stems from the divergent lifetime of certain phonon modes. Among two types of three-phonon processes, the difference process is forbidden at low temperatures, and the summation process cannot occur for some phonons in the lowest-lying branch due to the absence of decay channels [29]. These restrictions in the three-phonon process lead to the divergence of phonon lifetime at low temperatures. Some acoustic phonons of AlN and GaN with frequency < 10 THz and 7 THz, respectively, experience these restrictions below 100 K, and their lifetime and angular momenta increase drastically (Fig. 3).

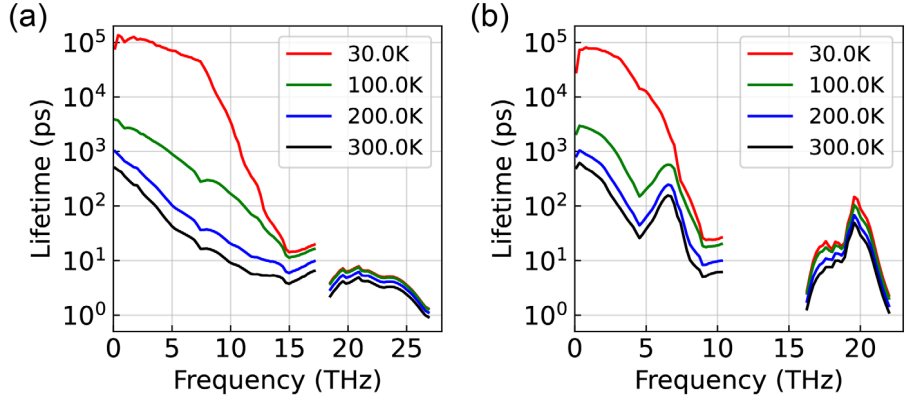

**Fig. 3. The phonon lifetime distribution of (a) AlN and (b) GaN.** Lifetime at each frequency is averaged over a frequency interval of ~0.13 THz.

At temperatures comparable to the Debye temperature, all phonons have finite lifetimes with relatively small variations, validating the constant lifetime approximation. At room temperature, which is about half (GaN) or a quarter (AlN) of the Debye temperature, $\alpha_{xy}$ is $1.03\times10^{-16}$ and $-2.90\times10^{-17}$ Js/m$^2$K with the average lifetime of 33.95 and 9.62 ps for GaN and AlN, respectively (Fig. 2a). It changes to $3.94\times10^{-16}$ and $8.61\times10^{-17}$ Js/m$^2$K with a proper consideration of mode-dependent lifetimes for GaN and AlN, respectively (Fig. 2b). Although the angular momentum of optical modes is dominant over the entire temperature range for both AlN and GaN within the constant lifetime approximation, the actual PAM at low temperatures is dominated by acoustic modes with divergent lifetime (Fig. 2). Especially for AlN, which has a high Debye temperature ~1150 K, PAM is mostly contributed by acoustic phonons at temperatures < 300 K. For GaN, the Debye temperature is relatively low ~600 K, and the optical phonon modes give a major contribution to PAM at temperatures > 50 K.

## B. g-BN

Monolayer g-BN has a graphene-like honeycomb structure but does not have inversion symmetry. Its point-group symmetry is $D_{3h}$ and the mode-specific PAM vector is always parallel to the out-of-plane direction ($z$-axis) (Fig. 4b). Among six phonon branches in g-BN, the ZA and ZO phonons have out-of-plane displacements and always have zero mode-specific PAM. Near the Γ point, the fourth phonon branch consists of ZO modes and has no angular momentum (Fig. 4b). Meanwhile, the fourth phonon branch at the K point possesses a large angular momentum of almost $\hbar$, which is the upper limit for the mode-specific PAM magnitude.

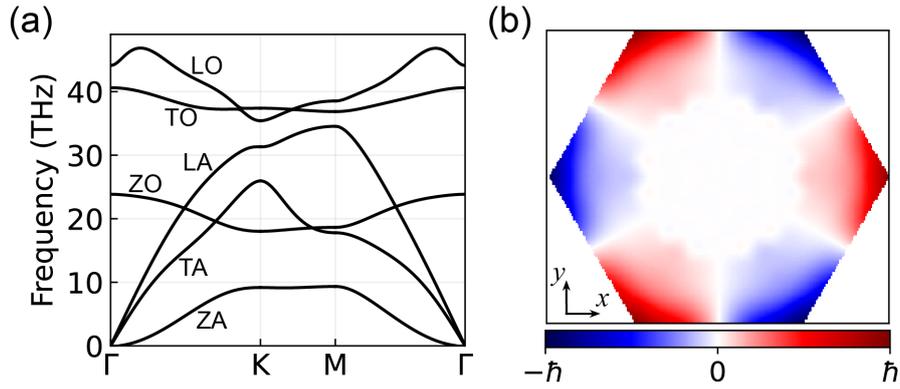

**Fig. 4. (a) Phonon band structure and (b) PAM texture of the fourth branch of g-BN.** The phonons at the K-point of the left panel in (a) are ZA, ZO, TA, LA, LO, and TO modes in ascending order in frequency.

The three-fold rotational symmetry of g-BN forces the PAM response tensor to be zero even in nonequilibrium conditions. In-plane uniaxial strain breaks the three-fold rotational symmetry and allows the phonon Edelstein effect to occur. The uniaxial strain along the zigzag or armchair direction changes the point group to $C_{2v}$ and produces the PAM response tensor with a non-zero component $\alpha_{zx}$. For example, the temperature gradient along the $x$-axis produces total PAM along the $z$-axis for both zigzag-directional (*x*-axis) and armchair-directional (*y*-axis) strains. The change in the harmonic phonon dispersion is studied for 5% tensile and compressive strains along the zigzag and armchair directions (Fig. 5). Under tensile strain, the in-plane optical modes (LO and TO) are mainly softened, and other changes are negligible. Under compressive strain, however, LO, TO, and LA modes are significantly hardened, and ZA modes are greatly softened, especially in direction of the applied strain.

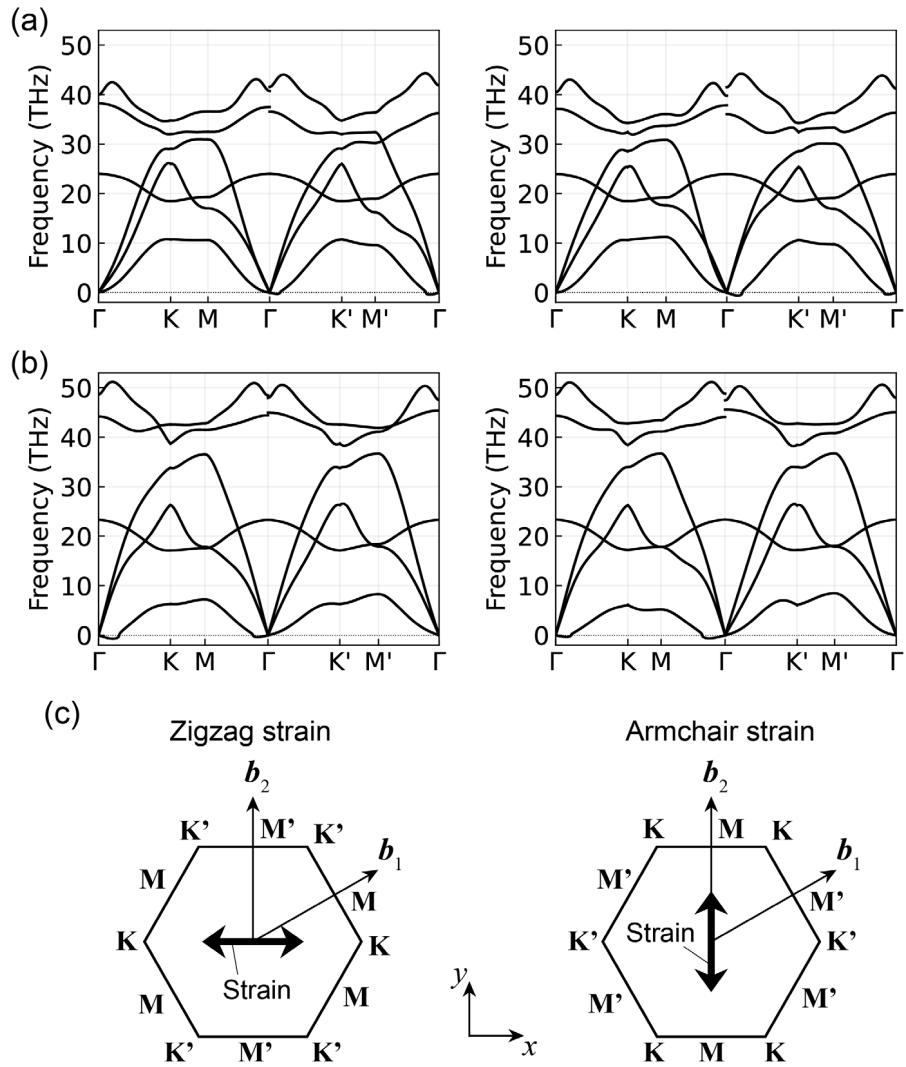

**Fig. 5. Phonon band structures of g-BN under strain.** (a) Tensile and (b) compressive strain of about 5% along the zigzag (left panel) and armchair directions (right panel). (c) Strain directions and high-symmetry points in the Brillouin zone.

The PAM response tensor was calculated with tensile and compressive strains applied (Fig. 6). In contract to wurtzite AlN and GaN, the PAM response tensor of g-BN with tensile strain did not diverge even near 0 K (Fig. 6). Similar to AlN and GaN, some phonons in the lowest-lying branch (ZA mode) in g-BN had divergent lifetimes at temperatures < ~50 K but in this case, they do not have angular momentum. This vanishing total PAM at low temperatures is a unique feature of perfectly-planar two-dimensional materials. In contrast, a compressive strain of ~5% produced a divergent PAM response tensor (Fig. 6b). At ~30 K, the PAM response tensor of g-BN under the compressive strain was ~150 times than under the tensile strain (–1.61×10$^{-25}$ and –1.11×10$^{-27}$ Js/mK for zigzag strain, and –2.63×10$^{-25}$ and –1.19×10$^{-26}$ Js/mK for armchair strain, respectively). The total PAM under compressive strain is the largest at 4 K and vanishes at 0 K since the term $\partial f_0/\partial T$ converges to zero strong enough to override the divergence of phonon lifetimes. In addition, we found out that the total PAM has opposite sign when the direction of strain is changed for temperature 100-300 K. This mechanism can be used to realize a device that inverts the PAM sign immediately by switching the strain between the *x*- and *y*-directions.

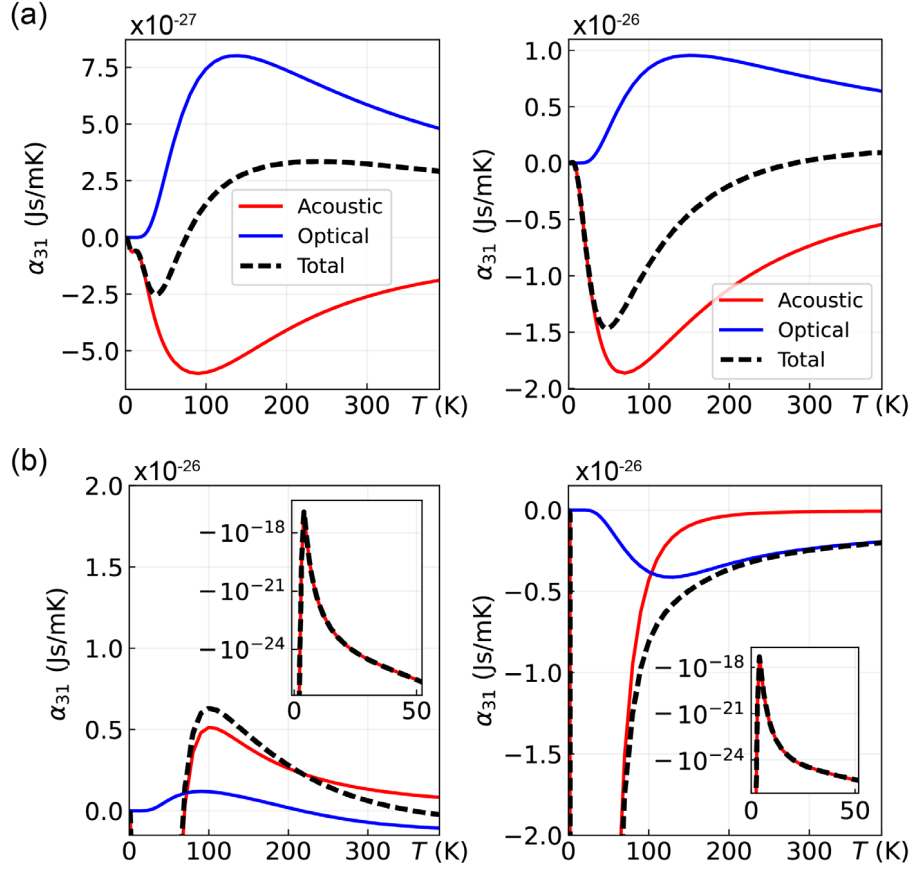

Fig. 6. $\alpha_{zx}$ of g-BN under (a) tensile and (b) compressive strain of 5% along the zigzag (left panel) and armchair directions (right panel). The acoustic (red), optical modes (blue) contribution, and the total (black dashed) are plotted. Insets in (b) are the logarithmic plot. The legends are the same as in Fig. 2.

Interestingly, the second-lowest lying branch (TA mode) caused the divergence of the PAM response tensor of the compressed g-BN. The ZA phonons softened in frequency by 22% on average under the compressive strain (Fig. 5). As a result, some phonons at ~15 THz in the TA branch have lost the decay channels for the three-phonon summation process, and their lifetime diverged (Fig. 7). TA phonons of compressed g-BN in the 13~18 THz range have a divergent lifetime of ~1 ns at 30K (Fig. 7b). One example of phonon triplets that gain significant lifetime by compressive strain is the summation process of three phonons with momenta $\mathbf{q}_1=0.317\mathbf{b}_1$, $\mathbf{q}_2=0.567\mathbf{b}_1-0.367\mathbf{b}_2$, and $\mathbf{q}_3=-0.25\mathbf{b}_1+0.367\mathbf{b}_2$,

where ($\mathbf{q}_1$, $\sigma$=2) phonon lies in TA branch and the other ($\mathbf{q}_2$, $\sigma$=1) and ($\mathbf{q}_3$, $\sigma$=1) phonons lie in ZA branch. Here, the momentum of phonons is conserved $\mathbf{q}_1 = \mathbf{q}_2 + \mathbf{q}_3$ since it is a normal process. The phonon frequencies are $\omega_{\mathbf{q}_1}^{TA}$=14.91, $\omega_{\mathbf{q}_2}^{ZA}$=9.05, and $\omega_{\mathbf{q}_3}^{ZA}$=5.81 THz without external strain, and $\omega_{\mathbf{q}_1}^{TA}$=14.92, $\omega_{\mathbf{q}_2}^{ZA}$=6.44, and $\omega_{\mathbf{q}_3}^{ZA}$=4.46 THz with 5% compressive strain along the zigzag direction. From the Fermi golden rule, such a three-phonon process under the compressive strain never occurs. As a result of the lack of decay channels, the lifetime of the phonon ($\mathbf{q}_1$, $\sigma$=2) increases greatly from 65.9 ps to 29.8 ns. Note that the phonon ($\mathbf{q}_1$, $\sigma$=2) still gains a finite lifetime through other decay channels.

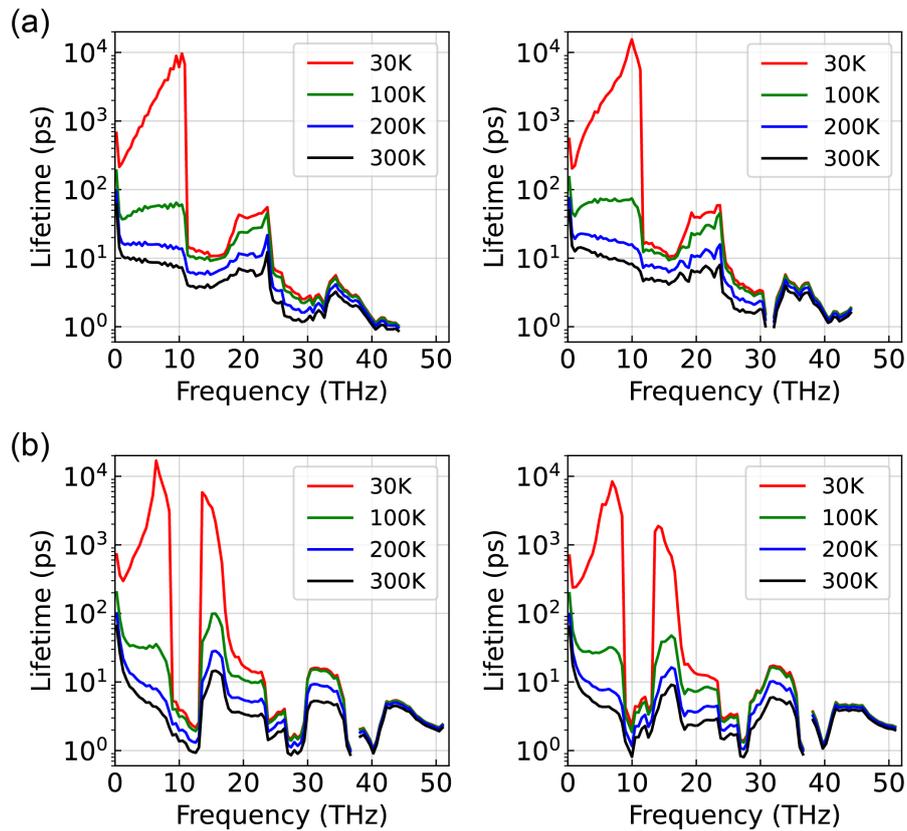

**Fig. 7. The phonon lifetime distribution of g-BN monolayer under (a) tensile and (b) compressive strain of 5% along the zigzag (left panel) and armchair directions (right panel) at four different temperatures.**

To identify the critical behavior of PAM of g-BN under compressive strain, we calculated the phonon lifetime and PAM with increasing strains (Fig. 8). $\alpha_{zx}$ increased by about two orders of magnitude at a critical strain of ~5 % (Fig. 8a). The ZA mode was gradually softened by the increase of compressive strain, and eventually, some specific TA-mode phonons lost the decay channels to ZA modes at the critical strain. The average phonon lifetime of g-BN at ~15 THz at 30 K increased by two orders of magnitude at the critical strain of ~5 % (Fig. 8b). This critical behavior of the PAM has interesting implication regarding its coupling to other angular momentum properties. The orbital motion of atoms with finite Born effective charge induces magnetization [27], and g-BN can exhibit a critical transition in magnetization as a result of divergent PAM that can be controlled by exerting compressive strain.

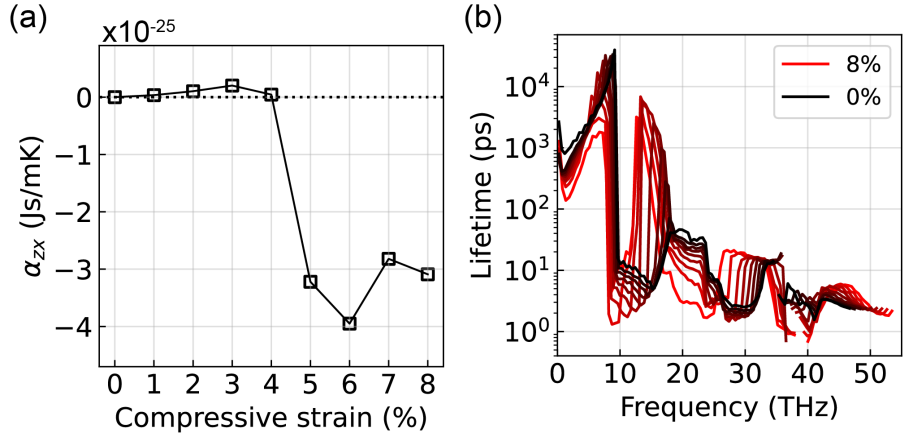

**Fig. 8. (a) $\alpha_{zx}$ of g-BN at 30 K under 0~8 % compressive strain along the zigzag direction and (b) their phonon lifetime.** The lines are colored in gradation from black to red in an interval of 1% strain.

The sign switching of PAM by strain in g-BN arises from the characteristics in the phonon band structure; the absence of the angular momentum for the modes in the lowest-lying branch and the blockage of the decay channels for the second lowest-lying branch due to the increase of frequency gap by the strain. If we can modify the dispersion of phonon modes, the range of temperature and the strain for the PAM switching can be engineered. To investigate such possibilities of PAM engineering, we constructed a dynamical matrix model for honeycomb structured materials including interatomic force constants up to the second-nearest neighbor interactions (see Supplemental Material [39] for the dynamical matrix model formalism and definitions of force constant parameters).

Our finding on the critical behavior of PAM implies that materials that possess very soft ZA modes may exhibit similar behavior in the PAM even without external strain. As tensile strain hardens the ZA modes, switching on and off the PAM can also be achieved by tensile strain, which is experimentally more feasible than compressive strain. Although the dynamical matrix model in Supplemental Material [39] does not directly present candidates for new PAM materials, it suggests a concept of designing PAM materials by phonon-band engineering.

## IV. DISCUSSION

The PAM and response tensor of wurtzite GaN, AlN, and monolayer g-BN were studied using first-principles calculations with full consideration of phonon lifetime. We found that the PAM response tensors of wurtzite GaN and AlN diverge at low temperatures due to the divergent phonon lifetime, which cannot be captured by the constant lifetime approximation. The g-BN monolayer can possess finite PAM by applying a strain that breaks its three-fold rotational symmetry. The PAM response tensor of g-BN exhibits a critical behavior at low temperatures under a compressive strain of about ~5% that blocks the decay channels for low-lying branches. However, tensile strains did not lead to divergence of the PAM response tensor. The possibility of PAM engineering was studied using a simple dynamical matrix model. The band width and gaps of phonon dispersions were found to be controllable by scaling the force constants and atomic mass ratio.

We note that our calculations did not consider how the lifetime is affected by the four-phonon process or by structural defects like isotopes. The actual lifetime of phonons should consider the additive property of phonon linewidth, i.e., $\tau^{-1} = \tau_{3-\text{ph}}^{-1} + \tau_{4-\text{ph}}^{-1} + \tau_{\text{iso}}^{-1}$. Our results of the divergent PAM may be overestimated if the contributions from other than the three-phonon process become dominant. However, four-phonon scattering events are rare for low-frequency phonons and at temperatures much lower than the Debye temperature [40, 41]. Heavier elements can be used to reduce overall phonon frequencies and suppress four-phonon processes together [40]. We note that the response tensor (Eq. 2) is calculated within harmonic phonon approximation since the thermal expansion of lattice constants for the materials in this study is very small (<~0.2% at 400K). Materials with relatively large thermal expansion may exhibit a significant shift in phonon frequencies at elevated temperatures. In such cases, the quasi-harmonic approximation and the temperature-dependent effective potential method [42] can improve the accuracy of the calculations.

The phonon Edelstein effect and its manipulation by temperature and strain in our study is a result of bosonic nature of phonons that enables the transport by low-lying phonon bands, which is in shear contrast to the electronic transport that is determined by the states near the Fermi level. Temperature-dependent PAM can also be utilized to the understanding of the thermal Hall effect in various materials such as 2D materials, van der Waals heterostructures, and superconductors. As PAM may couple with rotational excitations associated with charge, spin, or electromagnetic fields, our study demonstrates the potentials of mechanical or thermal engineering of such excitations in low dimensional materials.

## ACKNOWLEDGEMENT

This study was supported by the the National Research Foundation of Korea (NRF) grant (2022R1A2C1006530) funded by the Korea government (MSIT). Supercomputing resources including technical support was provided by Supercomputing Center, Korea Institute of Science and Technology

**103**, L100409 (2021).

**Supplemental Material for: Divergent phonon angular momentum driven by temperature and strain**

Young-Jae Choi and Seung-Hoon Jhi*

*Physics Department, POSTECH, Cheongam-ro 77, Pohang 37673, South Korea*

*Corresponding author: jhish@postech.ac.kr

**PAM textures of XNs (X=B, Al, and Ga)**

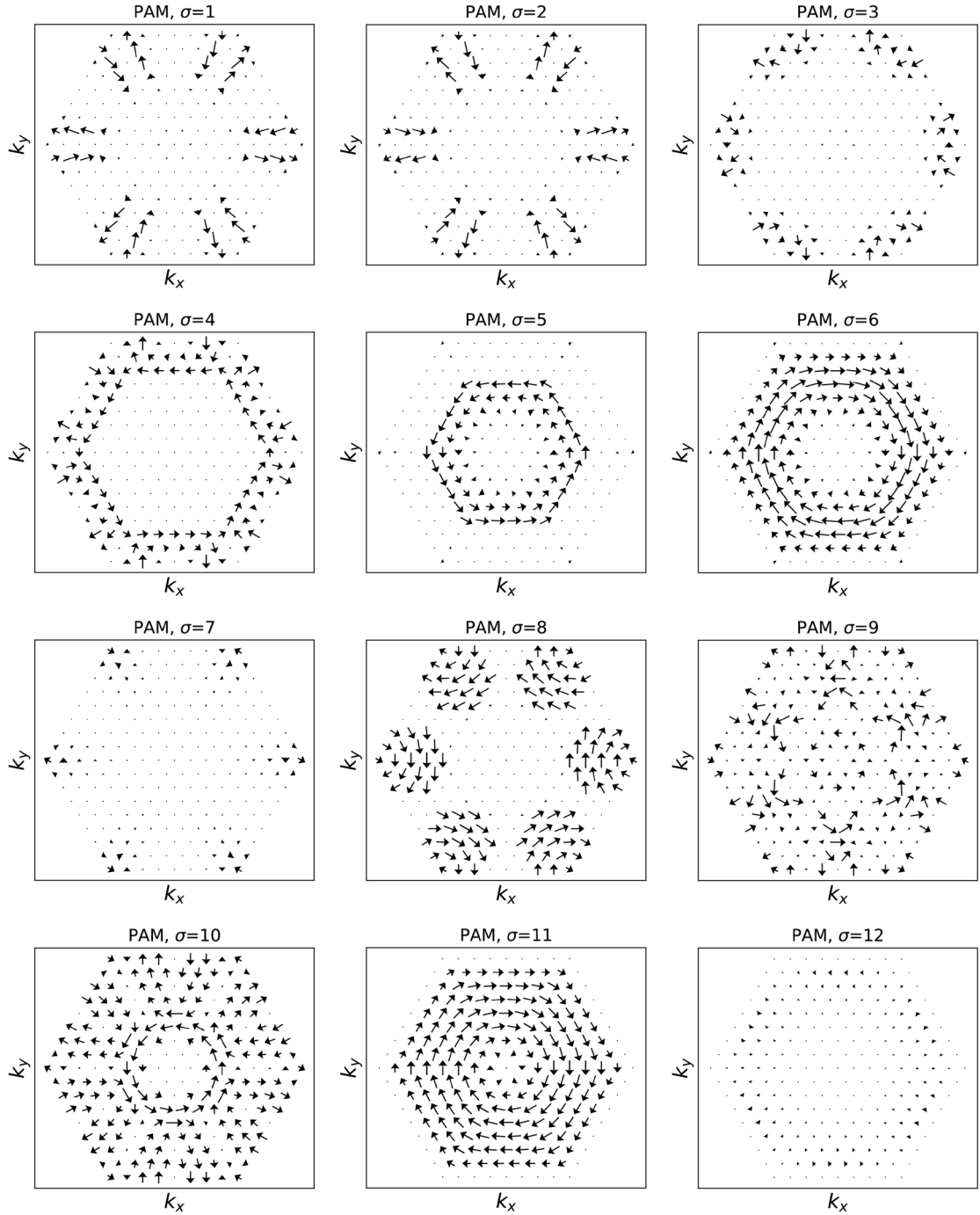

**Fig. S1. Mode-specific PAM textures of all phonon branches of wurtzite AlN.** $\sigma$ denotes the branch index of phonon bands.

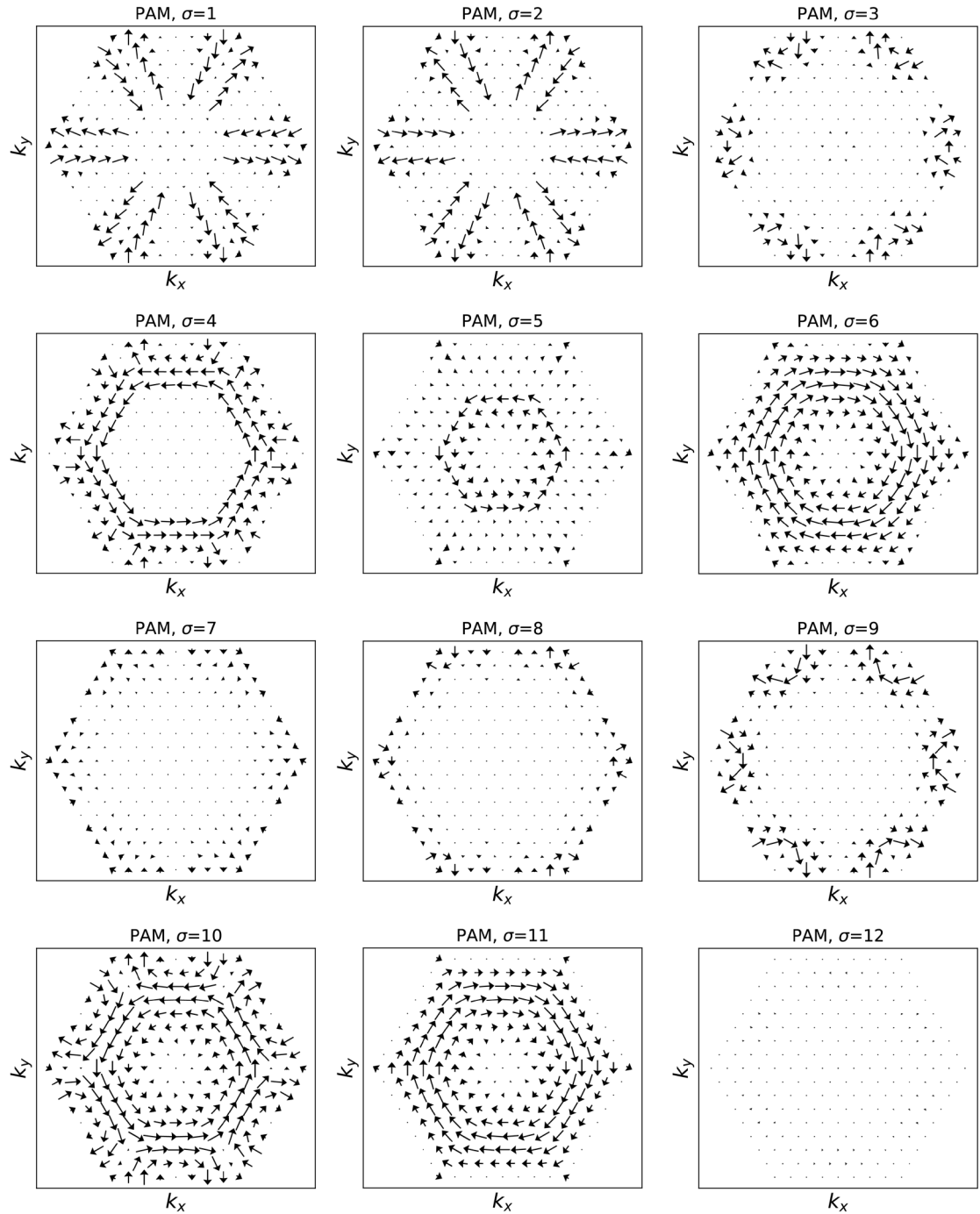

**Fig. S2. Mode-specific PAM textures of wurtzite GaN.**

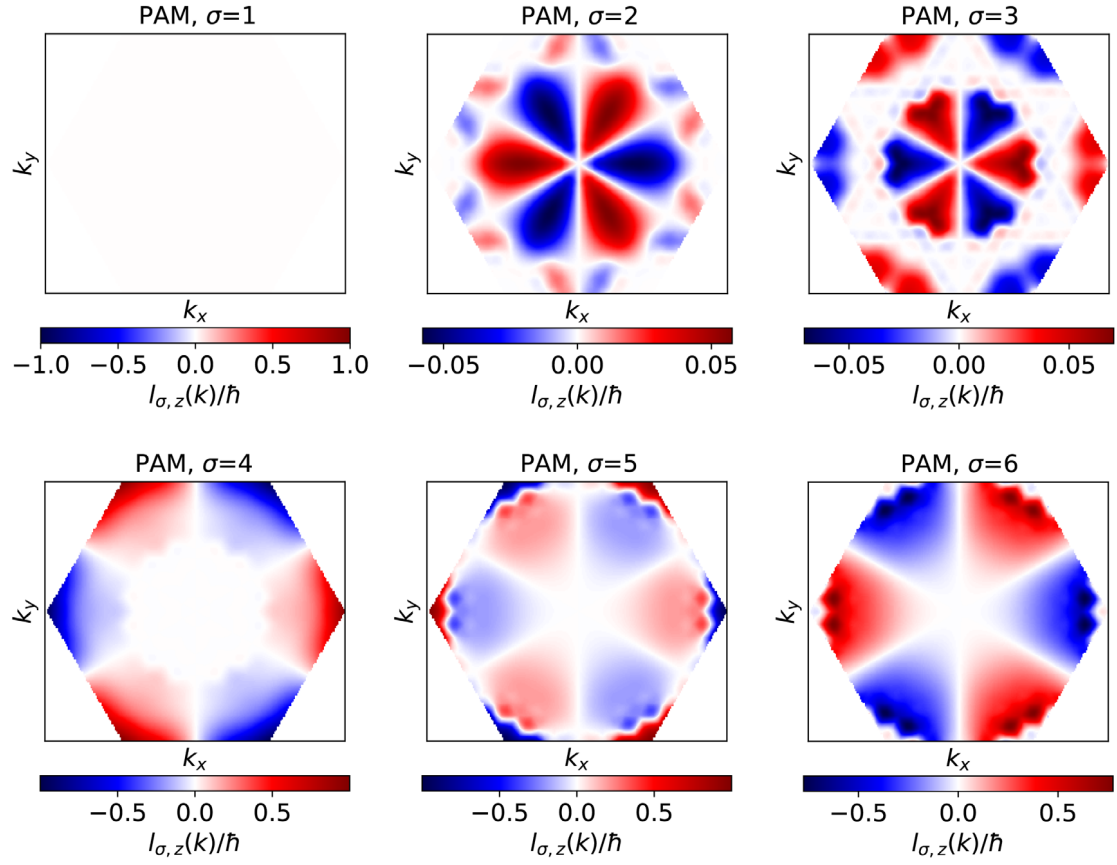

**Fig. S3. Mode-specific PAM textures of g-BN at equilibrium structure.** No PAM exists for $\sigma$=1.

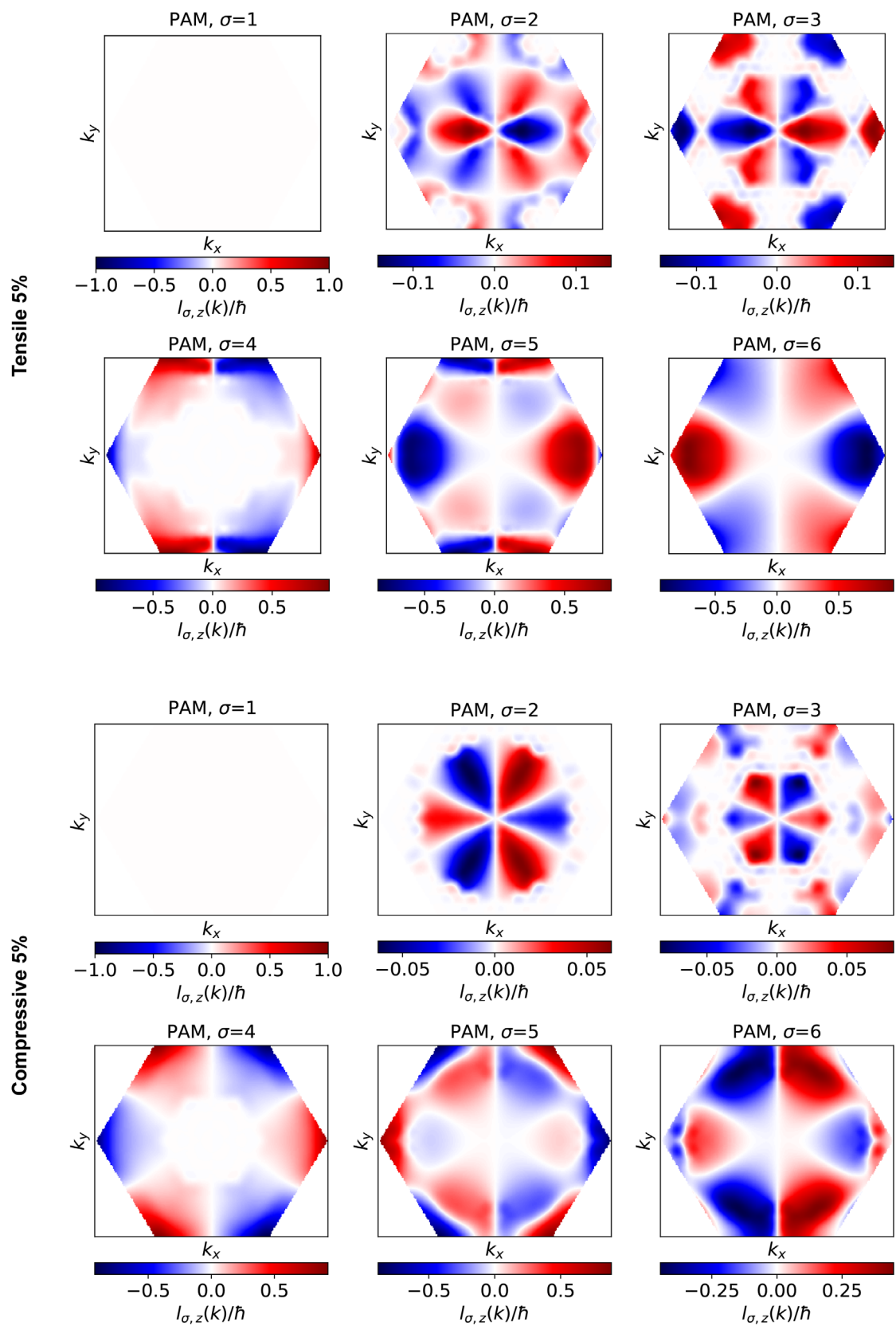

**Fig. S4. Mode-specific PAM textures of g-BN under tensile and compressive strains of 5% along the zigzag direction.** No PAM exists for $\sigma$=1.

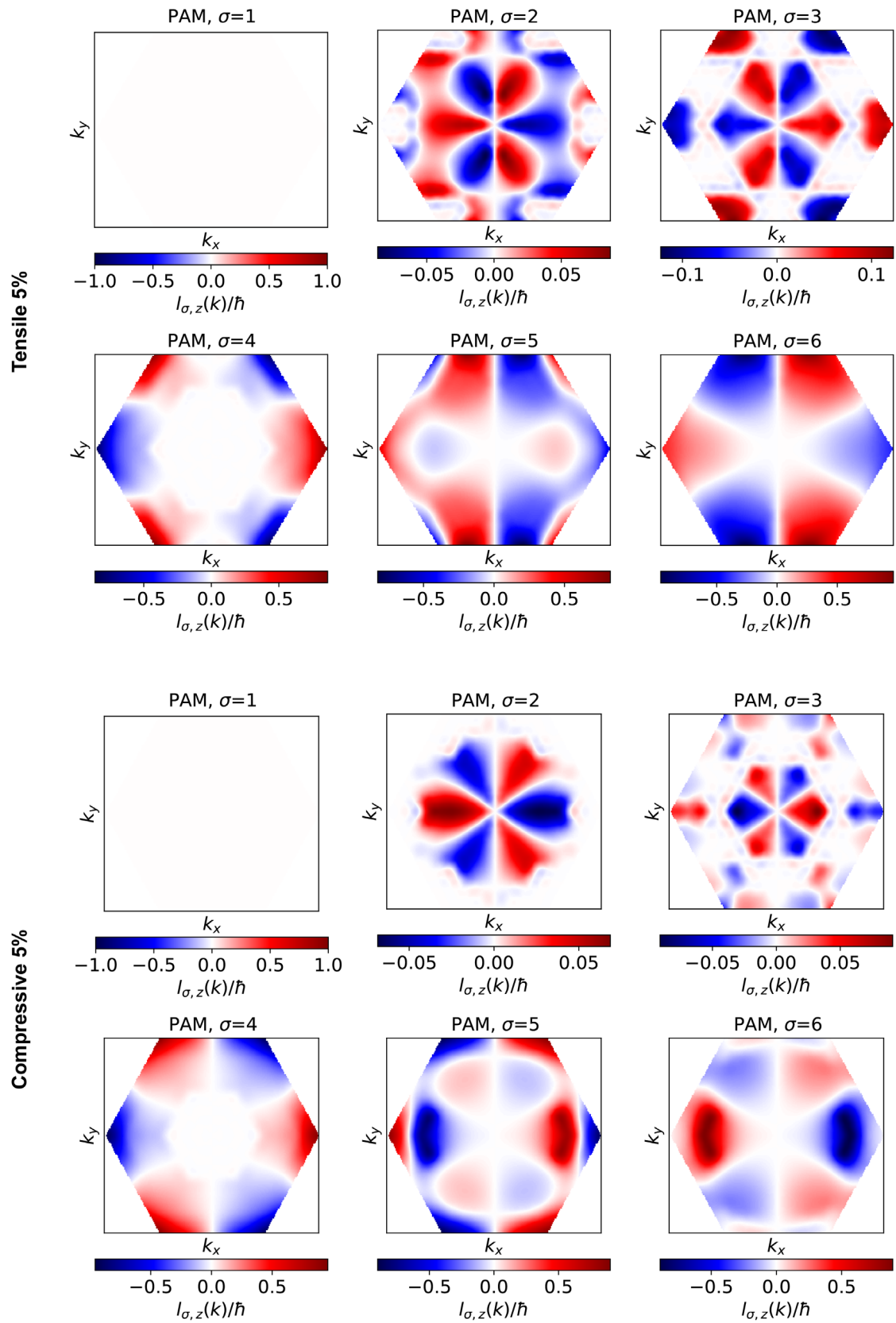

**Fig. S5. Mode-specific PAM textures of g-BN under tensile and compressive strains of 5% along the armchair direction.** No PAM exists for $\sigma$=1.

# Dynamical matrix model formalism for the g-BN structure

The dynamical matrix is expressed as

$$D_{jj'}^{\alpha\alpha'}(\mathbf{q}) = \frac{e^{i\mathbf{q}\cdot(\boldsymbol{\tau}_{j'}-\boldsymbol{\tau}_j)}}{\sqrt{m_j m_{j'}}} \sum_{\mathbf{R}'} A_{RR'jj'}^{\alpha\alpha'} e^{i\mathbf{q}\cdot(\mathbf{R}'-\mathbf{R})}, \tag{S1}$$

where $j$ the atomic index in the primitive unit cell, namely X(=B, Al, or Ga) or N, $\alpha$ the direction, $\mathbf{q}$ the crystal momentum, $\boldsymbol{\tau}_j$ the primitive cell position of the atom $j$, $m_j$ the atomic mass of the atom $j$, $\mathbf{R}$ the lattice vector, and $A_{RR'jj'}^{\alpha\alpha'}$ is the harmonic force constant. The analytical expressions for all dynamical matrix components up to the second-nearest neighbor force constants are expressed as

$$\begin{aligned}D_{jj}^{xx}(\mathbf{q}) = \frac{1}{m_j}\Big[&A_{\text{self},jj}^{\text{in}} \\ &+ 2\left\{\cos(\mathbf{q}\cdot\mathbf{a}_1) + \cos^2\frac{\pi}{3}[\cos\{\mathbf{q}\cdot(\mathbf{a}_1+\mathbf{a}_2)\} + \cos(\mathbf{q}\cdot\mathbf{a}_2)]\right\} A_{\text{2nd},jj}^{\text{LL}} \\ &+ 2\left\{\sin^2\frac{\pi}{3}[\cos\{\mathbf{q}\cdot(\mathbf{a}_1+\mathbf{a}_2)\} + \cos(\mathbf{q}\cdot\mathbf{a}_2)]\right\} A_{\text{2nd},jj}^{\text{TT}}\Big],\end{aligned} \tag{S2}$$

$$\begin{aligned}D_{jj}^{yy}(\mathbf{q}) = \frac{1}{m_j}\Big[&A_{\text{self},jj}^{\text{in}} + 2\left\{\sin^2\frac{\pi}{3}[\cos\{\mathbf{q}\cdot(\mathbf{a}_1+\mathbf{a}_2)\} + \cos(\mathbf{q}\cdot\mathbf{a}_2)]\right\} A_{\text{2nd},jj}^{\text{LL}} \\ &+ 2\left\{\cos(\mathbf{q}\cdot\mathbf{a}_1) + \cos^2\frac{\pi}{3}[\cos\{\mathbf{q}\cdot(\mathbf{a}_1+\mathbf{a}_2)\} + \cos(\mathbf{q}\cdot\mathbf{a}_2)]\right\} A_{\text{2nd},jj}^{\text{TT}}\Big],\end{aligned} \tag{S3}$$

$$\begin{aligned}D_{jj}^{xy}(\mathbf{q}) = \frac{1}{m_j}\Big[&2\sin\frac{\pi}{3}\cos\frac{\pi}{3}[\cos\{\mathbf{q}\cdot(\mathbf{a}_1+\mathbf{a}_2)\} - \cos(\mathbf{q}\cdot\mathbf{a}_2)](A_{\text{2nd},jj}^{\text{LL}} - A_{\text{2nd},jj}^{\text{TT}}) \\ &+ 2i\{\sin(\mathbf{q}\cdot\mathbf{a}_1) - \sin\{\mathbf{q}\cdot(\mathbf{a}_1+\mathbf{a}_2)\} + \sin(\mathbf{q}\cdot\mathbf{a}_2)\} A_{\text{2nd},jj}^{\text{LT}}\Big],\end{aligned} \tag{S4}$$

$$D_{jj}^{zz}(\mathbf{q}) = \frac{1}{m_j}\left[A_{\text{self},jj}^{\text{out}} + 2[\cos(\mathbf{q}\cdot\mathbf{a}_1) + \cos\{\mathbf{q}\cdot(\mathbf{a}_1+\mathbf{a}_2)\}] + \cos(\mathbf{q}\cdot\mathbf{a}_2)\, A_{\text{2nd},jj}^{zz}\right], \quad (S5)$$

$$D_{jj'}^{xx}(\mathbf{q}) = \frac{e^{i\mathbf{q}\cdot(\boldsymbol{\tau}_{j'}-\boldsymbol{\tau}_j)}}{\sqrt{m_j m_{j'}}}\left[\cos^2\frac{\pi}{6}(1+e^{-i\mathbf{q}\cdot\mathbf{a}_1})A_{\text{1st},jj'}^{\text{LL}} \right. \\ \left. + \left\{\sin^2\frac{\pi}{6}(1+e^{-i\mathbf{q}\cdot\mathbf{a}_1}) + e^{i\mathbf{q}\cdot\mathbf{a}_2}\right\}A_{\text{1st},jj'}^{\text{TT}}\right] \quad (j\neq j'), \quad (S6)$$

$$D_{jj'}^{yy}(\mathbf{q}) = \frac{e^{i\mathbf{q}\cdot(\boldsymbol{\tau}_{j'}-\boldsymbol{\tau}_j)}}{\sqrt{m_j m_{j'}}}\left[\left\{\sin^2\frac{\pi}{6}(1+e^{-i\mathbf{q}\cdot\mathbf{a}_1}) + e^{i\mathbf{q}\cdot\mathbf{a}_2}\right\}A_{\text{1st},jj'}^{\text{LL}} \right. \\ \left. + \cos^2\frac{\pi}{6}(1+e^{-i\mathbf{q}\cdot\mathbf{a}_1})A_{\text{1st},jj'}^{\text{TT}}\right] \quad (j\neq j'), \quad (S7)$$

$$D_{jj'}^{xy}(\mathbf{q}) = \frac{e^{i\mathbf{q}\cdot(\boldsymbol{\tau}_{j'}-\boldsymbol{\tau}_j)}}{\sqrt{m_j m_{j'}}}\sin\frac{\pi}{3}\cos\frac{\pi}{3}(-1+e^{-i\mathbf{q}\cdot\mathbf{a}_1})\left(A_{\text{1st},jj'}^{\text{LL}} - A_{\text{1st},jj'}^{\text{TT}}\right) \quad (j\neq j'), \quad (S8)$$

$$D_{jj'}^{yx}(\mathbf{q}) = D_{jj'}^{xy}(\mathbf{q}) \quad (j\neq j'), \text{ and} \quad (S9)$$

$$D_{jj'}^{zz}(\mathbf{q}) = \frac{e^{i\mathbf{q}\cdot(\boldsymbol{\tau}_{j'}-\boldsymbol{\tau}_j)}}{\sqrt{m_j m_{j'}}}(1+e^{-i\mathbf{q}\cdot\mathbf{a}_1}+e^{i\mathbf{q}\cdot\mathbf{a}_2})A_{\text{1st},jj'}^{zz} \quad (j\neq j'), \quad (S10)$$

where $\mathbf{a}_1$ and $\mathbf{a}_2$ two primitive lattice vectors, $A_{\text{self},jj}^{\text{in}}$ and $A_{\text{self},jj}^{\text{out}}$ the diagonal component of force constant matrix for in-plane and out-of-plane direction, respectively, $A_{\text{1st},jj'}^{\text{LL}}$, $A_{\text{1st},jj'}^{\text{TT}}$, and $A_{\text{1st},jj'}^{zz}$ the force constants between the first-nearest neighbors along the transverse, longitudinal, and z-axis directions, respectively, $A_{\text{2nd},jj'}^{\text{LL}}$, $A_{\text{2nd},jj'}^{\text{TT}}$, $A_{\text{2nd},jj'}^{\text{LT}}$, and $A_{\text{2nd},jj'}^{zz}$ are the force constant between the second nearest neighbors along each direction. The force constants between the in-plane and out-of-plane modes are all zero for those with perfectly planar 2-dimensional monolayer structures like graphene and g-XNs. The off-diagonal blocks of in-plane and out-of-plane blocks ($D_{jj'}^{xz}$, $D_{jj'}^{yz}$, $D_{jj'}^{zx}$, and $D_{jj'}^{zy}$) are zeros, and the 6x6 dynamical matrix of g-XN is block diagonal and decomposed into two square dynamical matrices for 4x4 in-plane and 2x2 out-of-plane modes, respectively.

# Band structures of g-XNs by dynamical matrix model

In order to validate the dynamical matrix model, its phonon band structures of three g-XNs are compared with first-principles calculations (Fig. S6). The first- and second-nearest-neighbor force constants are enough to reproduce phonon band structures of g-XNs overall. The acoustic sum rules are not enforced and we find the soft modes near Γ point due to this error. These artifacts can be removed by applying the translational and rotational sum-rules of Huang and Born-Huang to the force constant matrix [1, 2]. Other phonon modes are hardly affected by this procedure [2].

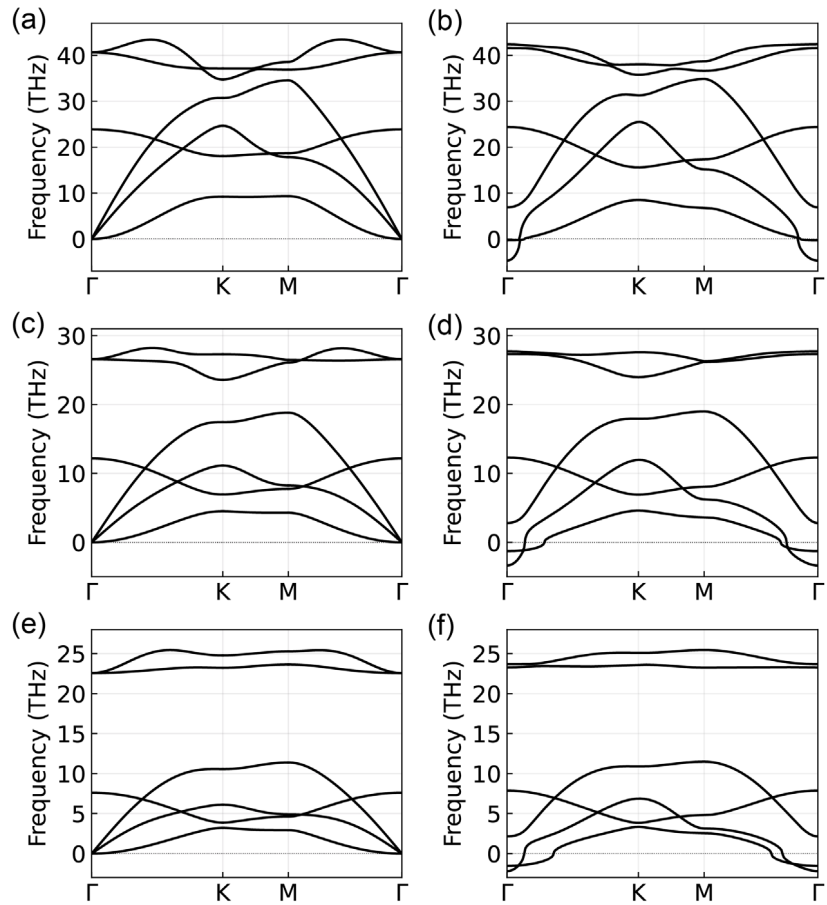

**Fig. S6. Harmonic phonon band dispersion of (a) g-BN, (b) g-AlN, and (c) g-GaN.** Left panel, first-principles calculations and right panel, the dynamical matrix model.

# Force constants of g-XNs

All independent components of the force constant matrix $A^{\alpha\alpha'}_{jj'RR'}$ of g-XNs up to the second nearest neighbor are listed in Table S1, where $A^{\alpha\alpha'}_{jj'RR'} = \partial^2 U/\partial\xi^{\alpha}_{jR}\partial\xi^{\alpha'}_{j'R'}$, $U$ the potential energy of the solid system, $\xi^{\alpha}_{jR}$ the displacement of $j$-th atom in $R$-th repeated unit cell along $\alpha$-direction ($x$, $y$, $z$, T (transverse), and L (longitudinal)). The direction T or L denote the displacement perpendicular or parallel to the line crossing the two atoms $(R, j)$ and $(R', j')$. First, there are four independent diagonal components of the force constant matrix: $A^{\text{in}}_{\text{self,XX}}$, $A^{\text{out}}_{\text{self,XX}}$, $A^{\text{in}}_{\text{self,NN}}$, and $A^{\text{out}}_{\text{self,NN}}$ for in-plane ("in") and out-of-plane direction ("out"). There are three independent first-nearest-neighbor force constants ($A^{\text{LL}}_{\text{1st,XN}}$, $A^{\text{TT}}_{\text{1st,XN}}$, and $A^{cc}_{\text{1st,XN}}$) and eight second-nearest-neighbor force constants. The remaining components are obtained from the antisymmetric matrix condition, $A^{\text{T}} = -A$.

**Table S1. Force constant and mass parameters for the dynamical matrix model.** Nitrogen mass $m_{\text{N}}$ equals 14.01 u (atomic mass unit).

|  | g-BN | g-AlN | g-GaN |
|---|---|---|---|
| $m_{\text{X}}$ (u) | 10.81 | 26.98 | 69.72 |
| $A^{\text{in}}_{\text{self,XX}}$ | 49.27 | 32.25 | 29.56 |
| $A^{\text{out}}_{\text{self,XX}}$ | 12.11 | 5.342 | 3.663 |
| $A^{\text{in}}_{\text{self,NN}}$ | 53.27 | 32.14 | 30.44 |
| $A^{\text{out}}_{\text{self,NN}}$ | 7.701 | 2.720 | 1.393 |
| $A^{\text{LL}}_{\text{1st,XN}}$ | -21.24 | -16.64 | -15.86 |
| $A^{\text{TT}}_{\text{1st,XN}}$ | -7.904 | -2.521 | -1.728 |
| $A^{zz}_{\text{1st,XN}}$ | -4.954 | -1.914 | -1.051 |
| $A^{\text{LL}}_{\text{2nd,XX}}$ | -4.664 | -2.488 | -2.542 |
| $A^{\text{TT}}_{\text{2nd,XX}}$ | 2.814 | 1.205 | 1.428 |
| $A^{zz}_{\text{2nd,XX}}$ | 0.4579 | 0.02102 | -0.1795 |
| $A^{\text{LT}}_{\text{2nd,XX}}$ | -0.9175 | -0.2863 | -0.5366 |
| $A^{\text{LL}}_{\text{2nd,NN}}$ | -4.912 | -2.045 | -2.016 |
| $A^{\text{TT}}_{\text{2nd,NN}}$ | 1.729 | 0.7054 | 0.4856 |
| $A^{zz}_{\text{2nd,NN}}$ | 1.193 | 0.5038 | 0.2520 |
| $A^{\text{LT}}_{\text{2nd,NN}}$ | -0.2871 | -0.1949 | -0.1133 |

## Phonon band engineering of g-BN model

The sign switching of PAM by strain in g-BN arises from the characteristics in the phonon band structure; the absence of the PAM for the modes in the lowest-lying branch and the blockage of the decay channels for the second lowest-lying branch due to the increase of frequency gap by the strain. If we can modify the dispersion of phonon modes, the range of temperature and the strain for the PAM switching can be engineered. To investigate such possibilities of PAM engineering, we constructed a dynamical matrix model for honeycomb structured materials including interatomic force constants up to the second-nearest neighbor interactions.

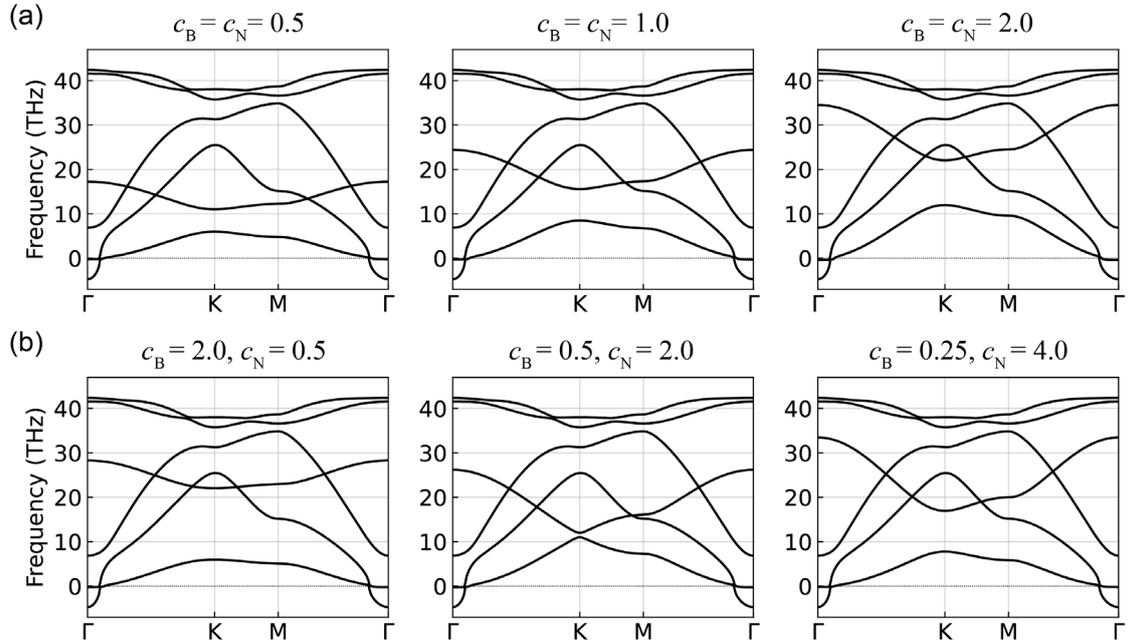

**Fig. S7. Phonon dispersion from the dynamical matrix model with out-of-plane force constants scaled by $c_B$ and $c_N$.** See Table S1 for the force constants of g-BN without scaling that correspond to the middle panel in (a).

The force constants of all out-of-plane modes ($A^{out}_{self,BB}$, $A^{out}_{self,NN}$, $A^{zz}_{1st,BN}$, $A^{zz}_{2nd,BB}$, and $A^{zz}_{2nd,NN}$) are scaled by $c_B$ and $c_N$ from the force constants of g-BN in nature. The force constants $A^{out}_{self,BB}$ and $A^{zz}_{2nd,BB}$ are multiplied by $c_B$, while $A^{out}_{self,NN}$ and $A^{zz}_{2nd,NN}$ are multiplied by $c_N$. The off-diagonal component $A^{zz}_{1st,BN}$ is multiplied by $\sqrt{c_B c_N}$ for consistency. Changing the out-of-plane force constants of a perfectly planar two-dimensional monolayered structure does not affect the in-plane modes. When symmetrically scaling the factors $c_B = c_N$, the frequency of the out-of-plane modes and the band gap between ZA and ZO branches were found to be proportional to the scaling constant $c_B$ and $c_N$ (Fig. S7a).

Then, the out-of-plane force constants of boron and nitrogen are scaled separately $c_B \neq c_N$ (Fig. S7b). When scaling $c_B$ and $c_N$, the product of two constants $c_B c_N$ is set to be 1 to preserve the ratio between the in-plane and out-of-plane force constants. For the atomic mass ($m_B$=10.81u, $m_N$=14.01u) and out-of-plane force constants ($A_{self,BB}^{out}$=12.11, $A_{self,NN}^{out}$=7.701) of boron and nitrogen in nature, the scaling constants of $c_B$=0.5 and $c_N$=2.0 produce the most graphene-like ZA- and ZO-mode dispersions (Fig. S7b, middle panel). In this case, the frequency gap between the ZA and ZO branches almost disappears. The ZA-ZO gap increases as the scaling parameters deviate from these values, which contributes to the blockage of decay channels for TA modes.

The critical behavior of PAM of g-BN under compressive strain results from the blockage of decay channels due to the increase in the gap between the ZA and TA branches. This finding implies that materials that possess very soft ZA modes may exhibit similar behavior in the PAM even without external strain. As the tensile strain hardens the ZA modes, switching on and off of the PAM can be achieved by the tensile strain as well, which is experimentally more feasible than the compressive strains. Although the force-constant model of this section does not directly present a candidate for a new PAM material search, it suggests a concept of PAM engineering based on the phonon band theories.

We tested the scaling of the cation-anion mass ratio, as well (Fig. S8). To compare with the band structures of g-XNs, the mass of cations is scaled to $2m_N$ and $5m_N$ comparable to the mass of Al and Ga, respectively. The resulting phonon bands are comparable to those of g-AlN and g-GaN in Fig. S6. The increment of average atomic mass is found to induce the softening of the in-plane acoustic modes (TA and LA). This finding is also true for the mass increment of the anion (nitrogen), but the ZA-ZO gap increases monotonically. This distinction attributes to the different magnitudes of force constants between cation and anion.

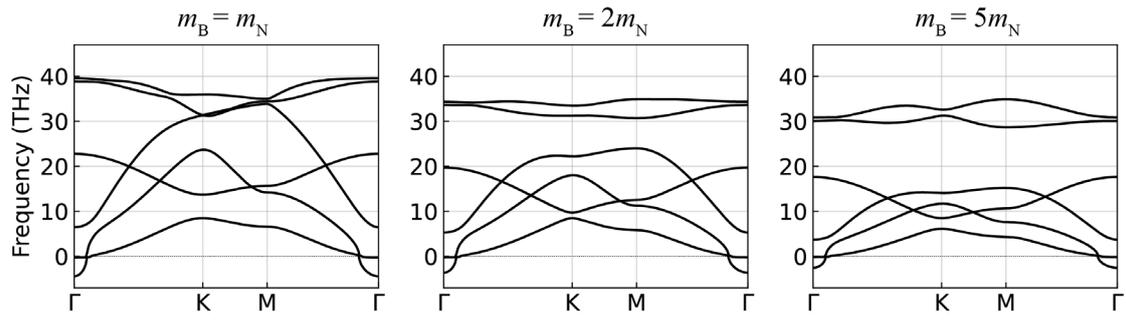

**Fig. S8.** Phonon dispersion from the dynamical matrix model by scaling mass of cation $m_B$.